\documentclass{emulateapj}
\def\gsim{\;\rlap{\lower 2.5pt
\hbox{$\sim$}}\raise 1.5pt\hbox{$>$}\;}
\def\lsim{\;\rlap{\lower 2.5pt
\hbox{$\sim$}}\raise 1.5pt\hbox{$<$}\;}

\def\etal{{\it et al.\thinspace}}
\def\eg{{\it e.g.\ }}

\newcommand{\be}{\begin{equation}}
\newcommand{\ba}{\begin{eqnarray}}
\newcommand{\ee}{\end{equation}}
\newcommand{\ea}{\end{eqnarray}}
\def\prosima{$\buildrel {\,\, \big \propto} \over \sim$}
\def\simpr{\lower0.5ex\hbox{\prosima}}

\begin{document}
\title{Identification of  a Fundamental Transition in a Turbulently-Supported Interstellar Medium}

\author{Evan Scannapieco, William J. Gray, \& Liubin Pan}
\affil{School of Earth and Space Exploration,  Arizona
  State University, P.O.  Box 871404, Tempe, AZ, 85287-1404.}

\begin{abstract}

The interstellar medium in star-forming galaxies is a multiphase gas in which turbulent support is at least as important as thermal pressure.  Sustaining this configuration requires continuous radiative cooling,  such that the overall average cooling rate matches the  decay rate  of turbulent energy into  the medium.   Here we carry out a set of numerical simulations of a stratified, turbulently stirred, radiatively cooled medium, which uncover a fundamental transition at a critical one-dimensional turbulent velocity of $\approx 35$ km/s. At turbulent velocities below  $\approx 35$ km/s,  corresponding to temperatures below $10^{5.5}$K, the medium is stable, as the time for gas to cool is roughly constant as a function of temperature.    On the other hand, at turbulent velocities above the critical value, the gas is shocked into an unstable regime in which the cooling time increases strongly with temperature, meaning that a substantial fraction of the interstellar medium is unable to cool on a turbulent dissipation timescale.   This naturally leads to runaway heating and ejection of gas from any stratified medium with a  one-dimensional turbulent velocity above $\approx 35$ km/s, a result that has  implications for galaxy evolution at all redshifts.\\

\end{abstract}


\section{Introduction}

The cosmic history of dark matter  is relatively simple.  Starting as weak  perturbations in a Gaussian random field, dark matter structures  accrete and merge to form ever-larger objects, leading to a hierarchical evolution that is in excellent agreement with a wide range of observations (\eg Fu \etal 2008; Vikhlinin \etal 2009; Komatsu \etal 2011). 

On the other hand, the formation of galaxies within these potentials is extremely complex.   Although gas initially collapses along with the dark matter,  it soon feels pressure support, cools radiatively,  forms molecules, and experiences a wide range of intricate physical processes.  Perhaps the most important of these is the formation of stars, which liberate vast quantities of energy that cause much of the gas to  reverse its history, parting ways from the dark matter and traveling outwards into the depths of intergalactic space. 

Such large-scale outflows have been observed in star-forming galaxies over a wide range of masses and redshifts (\eg Heckman, Armus \& Miley 1990;  Lehnert \& Heckman 1996;  Lehnert, Heckman, \& Weaver 1999; Martin 1999, 2005; Strickland \etal 2000, 2004; Pettini \etal  2001; Veilleux, Cecil  \& Bland-Hawthorn 2005; Westmoquette \etal 2007, 2009), and have been shown to be of crucial importance  in reconciling the distributions of galaxies and dark-matter halos (\eg Scannapieco \etal 2001; Benson \etal 2003). Yet, at the same time, galaxy outflows are very poorly-understood, as the interstellar medium (ISM) from which they form can not be modeled as a single-temperature gas.  Rather it is both continuously cooling and being stirred by turbulence, which in star forming regions is likely to be driven primarily by massive stars  (\eg McKee \& Ostriker 1977).  In fact, this simultaneous cooling and turbulent energy input is so extreme that it often leads to a supersonic medium in which turbulent velocities exceed the thermal velocities, and turbulent motions simultaneously support the disk and compress a fraction of the gas to drive star formation (\eg Elmegreen \& Scalo 2004;  Mac Low \& Klessen 2004).
      
The extremely short cooling times within the ISM make it impossible to model stellar feedback by simply adding thermal energy to the gas, but the range of physical scales involved does not allow for the direct simulation of supernova remnants within a galaxy-scale simulation. To deal with this problem, galaxy outflow simulations have suppressed cooling with an empirical heating function (Mac Low \etal 1989; Mac Low \& Ferrara 1999), imposed an artificial temperature floor (Tomisaka \& Bregman 1993; Tenorio-Tagle \& Mu\~ noz-Tu\~n\'on 1998; Suchkov \etal 1994; D'Ercole \& Brighenti 1999;   Strickland \& Stevens 2000;  Fujita \etal 2003; 2004),  delayed cooling by an arbitrary time delay  (Gerritsen \& Icke 1997; Thacker \& Couchman 2000), or simply added outflowing gas directly on large scales (Navarro \& White 1993; Mihos \& Hernquist 1994; Scannapieco \etal 2001; Springel \& Hernquist 2003).  While outflows arise from all of these approximations, the approaches are so simplified that  several studies have suggested that supernovae and stellar winds may not be effective at driving outflows at all, but rather the primary driver may instead be radiation pressure on dust (Thompson \etal 2005) or non-thermal pressure from cosmic rays (Socrates \etal  2008).

On the other hand, Scannapieco \& Br\" uggen (2010), showed that by using a subgrid model to track the unresolved velocities and length scales of turbulence driven by massive stars, one could produce galaxy-scale outflows while maintaining radiative cooling throughout the simulation.  Furthermore, the material ejected in these simulations was a multiphase distribution that grew out of the simultaneous turbulent heating and radiative cooling of the gas.   These results, although built on an approximate subgrid approach, illustrated that strong, stellar driven turbulence indeed had the potential to produce outflows with the properties of observed starbursting galaxies.

Here we examine this possibility more closely by conducting a series of direct numerical simulations that study the behavior of a stratified, turbulently supported medium, representing a small patch of a larger galaxy.    Previous theoretical work on stratified media at these scales  can be roughly grouped into two categories: (i)  two-and three dimensional studies that deposit the kinetic energy from a large number of supernovae within a single low-density superbubble (\eg Tomisaka \& Ikeuchi 1986; Mac Low \etal 1989; Tenorio-Tagle \etal 1990; Tomisaka 1998; Stil \etal 2009); and (ii) three-dimensional studies of the evolution of stratified turbulence driven by the stochastic injection of thermal and/or kinetic energy from individual or small groupings of supernovae
(\eg Korpi \etal 1999;  de Avillez \& Breitschwert 2004, 2005, 2007, 2010; Dib \etal 2006; Joung \& Mac Low 2006; Cooper \etal 2008; Shetty \& Ostriker 2008; Joung \etal 2009; Ostriker \& Shetty 2011).   
 
Here we adopt a third approach, abstracting the problem to study the long-term stability of a stratified medium that is continuously stirred at a fixed scale, and radiatively cooled such that the energy input from turbulent decay is matched to the overall cooling rate.   In this way our method is more like the ones adopted in studies of compressible turbulence in molecular clouds (\eg  Stone \etal 1998; Porter \etal 1999; Padoan \& Nordlund 1999; Cho \& Lazarian 2003; Cho \etal 2003; Padoan \etal 2004; Kritsuk \etal 2007; Benzi \etal 2008; Lemaster \& Stone 2008, 2009; Burkhart 2009), although unlike these studies we maintain vertical stratification of the medium, do not assume isothermal gas,  and include the atomic chemical reactions that are important in modeling the larger scale, higher temperature ISM.  While this more abstract approach limits the physical processes probed by our simulations, it nevertheless has three key advantages.

First, it sidesteps the many uncertain issues surrounding the coupling of massive stars with the surrounding medium.   In particular, the ISM simulations described above not only adopt many approximate techniques to deposit the energy from massive stars, but most of them also include an additional large source of diffuse heating, either taken from a purely numerical model that balances cooling in the initial configuration (\eg de Avillez \& Breitschwerdt 2004) or a model that approximates the photoelectric gas heating caused by the high-energy electrons that are removed from dust grains by stellar ultraviolet radiation (\eg Joung \& Mac Low 2006).   Our simulations, on the other hand, allow for cooling at all times in all locations, and they do not include any additional sources of energy beyond that of the turbulent stirring itself.

Secondly, our approach allows us to study the evolution of media with a wide variety of velocity dispersions without having to attempt to reproduce them by adjusting several indirect parameters such as diffuse heating rate, supernova rate, and the method used to deposit supernovae energy in the ISM.  Instead, as described below, we are able to increase the velocity dispersion simply by changing the mean density of the medium,  as this increases the overall average cooling rate, which, in equilibrium, is balanced by more vigorous turbulence.   Indeed, recent simulations that attempt to reproduce the turbulent ISM through direct supernova modeling find that the one-dimensional turbulent velocity dispersion changes only weakly with star formation rate, remaining $\lesssim 15$ km/s over a wide range of parameters (Dib \etal 2006; Shetty \& Ostriker 2008; Agertz \etal 2009; Joung \etal 2009).  On the other hand, many galaxies are observed to have velocity dispersions exceeding 30 km/s, and in some cases even exceeding 100 km/s (F\"orster Schreiber \etal 2006; 2009; Wright \etal 2007; van Starkenburg 2008; Epinat \etal 2008, 2009, 2010; Cresci 2009; Law \etal 2009; Genzel \etal 2008; 2011; Green \etal 2010; Jones \etal 2010; Lemoine-Busserolle \& Lamareille 2010; Lemoine-Busserolle \etal 2010).

Finally, and perhaps most importantly, our approach lets us separate effects arising from the direct deposition of hot gas onto the grid from those caused purely by the presence of widespread turbulence. This in turn allows us to identify for the first time a global instability that occurs in turbulently-supported media driven above a critical velocity dispersion.

The structure of this work is as follows.  In \S 2, we describe our numerical simulations of stratified, turbulently-stirred, radiatively-cooled media. In \S3 we present the results of these simulations, focusing on the physics of a transition that occurs at a critical turbulent velocity dispersion.    A  discussion is given in \S4, which also explores the implications of our results for galaxy outflows across cosmic time. 

\section{Methods}

To study the stability of turbulently-supported media as a function of velocity dispersion, we carried out a suite of simulations using  FLASH (version 3.2), a multidimensional hydrodynamics code (Fryxell \etal  2000) that solves the Riemann problem on a Cartesian grid with a directionally-split  Piecewise-Parabolic Method (PPM) (Colella \& Woodward 1984; Colella \& Glaz 1985; Fryxell, M\" uller, \& Arnett 1989).  While FLASH is capable of adaptive mesh refinement  calculations,  in each of our simulations,  the hydrodynamic  equations were solved on  a fixed  grid, in a box of size $2 H$ in the $x$ and $y$ directions, spanning $-4 H$ to $4 H$ in the $z$ direction, where $H$ is the gravitational scale height  of the gas distribution.  In our fiducial simulations we adopted a $128\times128 \times 512$ grid such that $H/\Delta x = 64.$  The boundary conditions were taken to be periodic in both the $x$ and and $y$ directions, and in the $z$ direction we adopted the FLASH ``diode" boundary condition, which, like the outflow boundary condition, assumes a zero normal gradient for all flow variables except pressure, but unlike the  outflow boundary condition, does not allow material to flow back onto the grid.

In the $z$ direction we also assumed a gravitational acceleration given by 
\be
g=g_0 \frac{-z}{\sqrt{z^2+a^2}},
\label{eq:g}
\ee
which contains a smooth transition at $z=0$, but  is nearly constant outside of $|z|=a$, where we fixed $a=H/2.$
For our initial conditions we adopted a constant temperature of $T_{\rm init} = 2 \times 10^5$K throughout the simulation
and a density distribution given by
\be
\rho(z,t) = \rho_0 e^{-\left[ \frac{(z^2+a^2)^{1/2}-a}{H}\right]}.
\ee  
If it were to cool only moderately to $10^5$K, this distribution would remain in hydrostatic equilibrium if  $g_0 = k 10^5 {\rm K} /(\mu m_p H)$ where $k$ is the Boltzmann constant, $m_p$ the mass of the proton and for ionized gas $\mu=0.6$, and we adopted this value of $g_0$ as our fiducial one. However, in cases in which the medium is cooled and stirred vigorously different values of $g_0$ may be appropriate, and we also explored the impact of varying this value, expressing it in terms of $T_{\rm grav}$ such that $g_0 = k T_{\rm grav} /(\mu m_p H).$  This means that, if it is not slowed down by material above it, gas will be able to escape from the midplane to the simulation boundaries at $\pm 4H$ if its velocity exceeds $v_{\rm esc} = \sqrt{2 g_0 (4H)} = 108$ km/s $\sqrt{T_{\rm grav}/10^5 {\rm K}}.$

 Our simulations included seven species: H$^+$, H, H$^-$, He, He$^+$, He$^{++}$, and electrons.
The gas was assumed to be initially 76\% ionized hydrogen and 24\% singly ionized helium by mass,  but this mix was able to change in time as we implemented turbulent velocity forcing, cooling, and atomic chemistry, as described in more detail below.  

\subsection{Velocity Forcing}

The fluid equations solved in our simulations are
\ba
\frac{D{\rho}}{D t} &=& 0, \label{eq:rho} \\
\frac{D \rho u_i}{D t} +\frac{\partial
P}{\partial x_i} &=&  \rho g_i + \rho f_{i}  F_{\rm cold}  e^{-\left[ \frac{(z^2+a^2)^{1/2}-a}{H}\right]},
\label{eq:u}\\
\frac{D \rho E}{D t} 
+\frac{\partial P  u_j}{\partial x_j}
&=&    \rho \dot E_{\rm cool}  +  \rho \dot E_{\rm chem}, \\
\frac{D{ \rho} X_s}{D t}  
&=& \rho A_s \dot R_s  \label{eq:species}.\\
\nonumber
\ea
On the left hand side of these equations
$\rho({\bf x},t)$ is the  density field,  $u_i({\bf
x},t)$  is the velocity field in the $i$
direction, $P({\bf x},t)$ is the pressure,  $E({\bf x},t)$ is
the total internal energy, and $X_s({\bf x},t)$ is the mass fraction of species
$s$, where  $t$ and ${\bf x}$ are time and position variables, and 
$\frac{D}{D t} \equiv \frac{\partial}{\partial t} + \frac{\partial u_j}{\partial x_j}.$
On the right hand side $g_i$ is the acceleration of gravity, ${\bf f}({\bf x},t)$ is a turbulent driving force, $F_{\rm cold}$ is the fraction of $T \leq 1.5 \times 10^4$ gas, $\rho \dot E_{\rm cool}$ is the energy loss due to radiative losses,  $\rho \dot E_{\rm chem}$ is the change in internal energy due to chemical reactions, and   $\rho A_s \dot R_s$ is the change in the mass fraction of species $s$ due to ionizations and recombinations.

As our goal is to study the evolution of a cooling ISM that is purely heated by the decay of turbulence, no hot gas was added directly to the simulations at any time. Rather we perturbed the velocity field  using  the FLASH3 ``Stir'' package (Benzi et al.\ 2008) which  we modified as described in Pan \& Scannapieco (2010).    To cleanly differentiate between the impact of turbulence from other forms of energy input,  we avoided adding $P dV$ energy to the system directly through our forcing term, and assumed that the flow driving terms, ${\bf f}({\bf x},t),$ contained only solenoidal  modes, i.e., $\nabla \cdot {\bf f} =0.$
Note however, Pan \& Scannapieco (2010)  found that the energy dissipation timescale in supersonic turbulence is actually 
independent of the fraction of driving energy contained  in compressible modes. 

We took $\bf{f}$ to be a Gaussian random vector with an exponential temporal  correlation.  This forcing scheme with a finite correlation timescale is different from that in studies using an  infinitesimal correlation timescale with independent forcing at each  time step (\eg Lemaster \& Stone 2008), or an infinite correlation  timescale with a fixed driving force (\eg Kritsuk et al.\ 2007).  
However,  the choice for the value of $t_f$ is not very important  because it does not affect the energy dissipation timescale  (Pan \& Scannapieco 2010). 

As in Pan \& Scannapieco (2010) the behavior of the $f_i$ term can be summarized as  $\langle f_i({\bf k}, t)f_j({\bf k}, t') \rangle = \mathcal{P}_{\rm f}(k)  \left( \delta_{ij} - \frac{k_i   k_j}{k^2}\right) \exp \left[-\frac{(t-t')}{t_{\rm f}} \right],$ where $t_{\rm {f} }$ is the forcing correlation timescale, taken to be $20\%$ of $H$ over the initial sound speed.
  Unlike Pan \& Scannapieco (2010), however, $f_i$  was   multiplied by two important factors.   The first factor,
  $F_{\rm cold}(t),$ is the fraction of cold, $T \leq 1.5 \times 10^4$K  gas in our simulation as a function of time.  This
 acted to increase the level of stirring as gas cooled, causing a feedback loop that results in an equillibrium configuration with $F_{\rm cool}$ roughly fixed.  For each simulation we varied the magnitude of $f_i$ such that this fraction was $\approx 0.1$ over the initial quasi-stable phase discussed below.  This approach is meant to approximate the situation in a real galaxy in which excessive cooling would lead to rapid star formation and energy input. Furthermore, the second term, $ \exp{\left\{- [(z^2+a^2)^{1/2}-a] H^{-1} \right\}}$, causes stirring to be most vigorous in regions with higher initial densities.  The inclusion of this term   approximates the concentration of energy input from stars near their sites of formation near the midplane of their host galaxies.
  
Also unlike in Pan \& Scannapieco (2010), the forcing wave  numbers were chosen to be in the range $10 \le H k/2\pi \le  10.25.$
To characterize the large scale properties of the flow, we defined a forcing length scale, $L_{\rm f}$, from the forcing spectrum,  $L_{\rm   f} = \int  \frac{2 \pi} {k} \mathcal{P}_f(k) d{\bf k}/\int \mathcal{P}_f(k) d {\bf k} = H/10.$  This length scale is chosen such that the turbulent driving occurs on scales substantially smaller than the vertical scale height, as would be the case in a real galaxy.  In fact, as discussed further below, even smaller driving scales would be expected in nature, but these are not possible in our current simulations because they would require adding velocity kicks on unresolved scales.

As a measure of the overall level of turbulence at each timestep, we calculated the volume-weighted average one-dimensional velocity dispersion of the material as
\be
\sigma^2_{\rm 1D,V}(t) = \frac{ \sum_{i=1}^N {\bf u}({\bf x_i},t)^2 }{3 N},
\label{eq:sigV}
\ee
where $i$ is a counter over all $N$ cells in the simulation, and the factor of 3 accounts for the ratio of  the  observable velocity dispersion in a single direction $\sigma_{\rm 1D}$, from the unobservable total velocity dispersion in all three directions $\sigma_{\rm 3D}.$ As emphasized in Joung, Mac Low, \& Bryan (2009) the velocity dispersion is really a scale-dependent quantity
\be
\sigma^2_{\rm 1D,V}(t,r) =  \frac{1}{4 \pi V } \int d{\bf x} \int d \Omega \left\{ [{\bf u}({\bf x},t)-{\bf u}({\bf x} + {\bf r},t)] \cdot {\bf \hat r} \right\}^2,
\ee
where the first integral is over the full simulation volume, $V,$  and the second integral averages the longitudinal velocity difference over all point separated by a distance $r.$ However we find that at all scales greater than or equal to the forcing scale $\sigma^2_{\rm 1D,V}(t,r)$ agrees with $\sigma^2_{\rm 1D,V}(t)$ within $\approx 10\%$ and eq.\ (\ref{eq:sigV}) represents a robust value that accurately  characterizes the overall  properties of the flow.
Thus we used a similar method to calculate the mass averaged one-dimensional velocity dispersion as
\be
\sigma^2_{\rm  1D,M} (t) = \frac{\sum_{i=1}^N \rho({\bf x_i},t){\bf u}({\bf x_i},t)^2 }{3 \sum_i^N \rho({\bf x_i},t)},
\ee
and the volume-weighted average 1D velocity dispersion as a function of height
\be
\sigma^2_{\rm  1D,V} (t,z) = \frac{\sum_{j=1}^J {\bf u}(\vec x_j,z,t)^2 }{3 J},
\ee
where $j$ is now a counter over all $J$ cells at a given height $z$.

In each of our simulations, we find that the volume-weighted average rms turbulent velocity as a function of  height  scales with the density profile as $n(z)^{1/2}$. This can be explained because the height dependence of the driving  force is chosen to be proportional to the density, such that  the energy input rate per unit mass goes like $\propto f^2 n(z)^2 L_f/\sigma_{1D,V}(z,t),$ where the large eddy turnover, $L/\sigma_{\rm 1D,V},$ enters because it is essentially the timescale over which the driving forcing is correlated. Equating the energy input rate to the turbulent energy  dissipation rate $\approx \sigma_{1D,V}^3/L_f$ then gives $\sigma_{1D}(z,t) \propto n^{1/2}$, which is biased towards the midplane where turbulent driving would be the most vigorous in a real galaxy 

Finally, we note that McCourt \etal (2011) and  Sharma \etal (2011) have identified that the behavior of vertically stratified medium simultaneously experiencing optically-thin cooling and uniform heating only develops a multiphase distribution if the  the free-fall timescale  $t_{\rm ff}= \sqrt{H/g}$ is much longer than the overall average cooling time (where the local cooling time is defined by eq.\ \ref{eq:tcool} below). As our simulations are carried out in an equillibrium configuration in which the average cooling time, $\bar t_{\rm cool}$ is  approximately equal to the turbulent dissipation timescale, $t_{\rm diss} = L_f/\sigma_{\rm 3D,M},$ and 
\be
\frac{t_{\rm ff}}{\bar t_{\rm cool}}  \approx \frac{H}{L_f} \frac{10^5 K}{T_{\rm grav}} \frac{\sigma_{1 D,M}}{22 {\rm \, km/s}},
\ee
is well above one in all of our runs.   Note however, that while the average cooling time is always much less than
$t_{\rm ff},$ the local cooling time can vary significantly from region to region due to spatial variations of the flow density 
and temperature. Furthermore, the energy dissipation in a turbulent flow exhibits strong 
spatial fluctuations, a phenomenon know as intermittency.  This leads to nonuniform heating 
rate by turbulent energy dissipation (Pan \&  Padoan 2009, Pan et al.\ 2009), which gives rise to 
the spatial variations of the temperature.  In fact, as we shall see in more detail below, it is the variance in energy dissipation and cooling times that lead to a global instability.

\subsection{Cooling} 

The energy input from turbulent driving in our simulations,  as in a real galaxy,  
is balanced by radiative cooling. This is implemented  in the optically-thin limit, 
assuming local thermodynamic equilibrium as   
\be
\dot E_{\rm cool}  =  (1-Y)(1-Y/2) \frac{\rho \Lambda(T,Z)}{(\mu m_p)^2},
\label{eq:ecool}
\ee 
where $\dot E_{\rm cool}$ is the radiated energy per unit mass,
$\rho$ is the density in the cell, $m_p$ is the proton mass, $Y$ is
the helium mass fraction, $\mu$ the mean atomic mass, and
$\Lambda(T,Z)$ is the cooling rate as a function of temperature and
metallicity.     Here we made use of the tables compiled by  Wiersma,
Schaye, \& Smith (2009) from the code CLOUDY (Ferland \etal 1998),
making the simplifying approximations that the metallicity of the gas is always solar and that the
abundance ratios of the metals always occur in solar proportions.   As in Gray \& Scannapieco (2010), subcycling was
implemented  within the cooling routine itself, such that $T$ and $\Lambda(T,Z)$  were recalculated every time $E_{\rm cool}/E > 0.1$.  This is equivalent to an integral formalism that assumes a constant density over each hydrodynamic time step (Thomas \& Couchman 1992; Scannapieco, Thacker, \& Davis 2001).  In order to give the distribution a chance to attain an initial level of turbulence, turbulence is stirred as if $F_{\rm cool} = 0.1$ and cooling is shut off for the first 0.25 $H$/(50 km/s) of the simulation. 

\begin{figure}
\epsscale{1.2}
\plotone{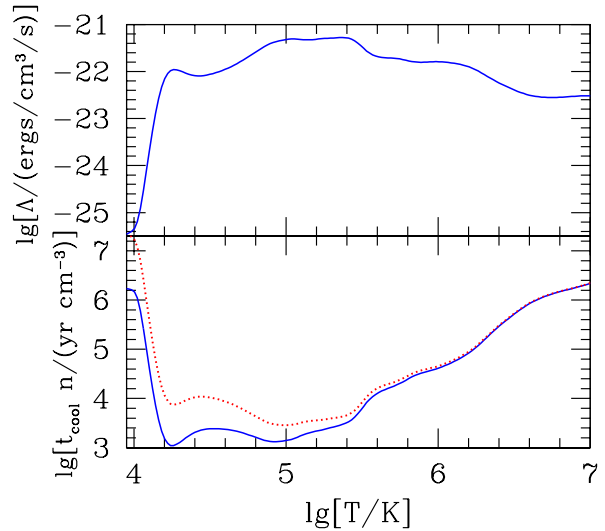}  
\caption{{\em Top:}  Radiative cooling rate as a function of temperature for solar metallicity gas, as is used in our simulations.  {\em Bottom:} Cooling time of the gas as a function of temperature. The solid line shows the cooling time as defined as $1.5 n {\rm k}T/(\rho \dot E_{cool})$, and the dotted curve shows the impact of recombinations on the cooling time as estimated by $(1.5 n{\rm k}T+ n 13.6eV)/(\rho \dot E_{cool}).$}
\label{fig:cool}
\end{figure}

In the upper panel of Figure \ref{fig:cool}, we show the radiative cooling rates used in our models. Here the peak at $T \approx 2\times10^5$K is dominated by  line emission from metal ions, whose atomic energy levels are easily excited by collisions at these temperatures.   Above this temperature, many atoms become fully ionized and the effectiveness of line cooling decreases, while below this temperature collisions become too weak to excite most atomic transitions.   Finally, the strong drop below about $15,000$K corresponds to the point at which even the high-energy tail of the kinetic energy distribution is unable to excite atomic hydrogen transitions.  Below this level, cooling by dust emission and molecular transitions become important, but these are much less efficient and not considered in our simulations, which are focused on the evolution of atomic gas.

In the lower panel of this figure, we plot the radiative cooling time, defined as the energy per unit volume  divided by the radiative cooling rate
per unit volume 
\be
t_{\rm cool} \equiv 1.5 n {\rm k} T/(\rho \dot E_{\rm cool}),
\label{eq:tcool}
\ee
which is inversely proportional to density.  From $1.5\times 10^4$ to $\approx 2 \times 10^5$K, $\Lambda$ is approximately proportional to $T$ such that the cooling time is roughly constant as a function of temperature.  Above $2 \times 10^5$K, on the other hand $\Lambda$ drops strongly with temperature, meaning that gas that is heated to these temperatures takes orders of magnitude longer to cool.  Finally, also shown on this plot is an estimate of the cooling time that approximately takes into account the additional energy released as gas moves from ionized to neutral.   This shows that this additional energy makes a substantial difference in the evolution of the gas below $2\times 10^5$K, an effect we capture in detail by taking account of atomic chemistry throughout our simulations.

 \subsection{Atomic Chemistry}
 
To handle ISM ionizations and recombinations, we used a modified version of the chemistry solver described in Gray \& Scannapieco (2010).
As in that study, we defined the molar abundance of the $i$th species as, $Y_s \equiv X_s / A_s,$  where, as in eq.\ (\ref{eq:species}) $A_s$ is the atomic number and  $X_s$ is the mass fraction of species $s,$ such that conservation of mass is given by $\sum^N_s X_s = 1.$ The chemical evolution of each the 7 species  in our simulations was cast  as  a continuity equation in the form 
 \begin{equation}
\frac{dY_s}{dt}  =  \dot{R_s},
 \end{equation}
where $\dot{R_s}$ is the total reaction rate due to all the binary reactions of the form $i+j \rightarrow s+l,$ defined as
\begin{equation}
\dot{R_s} \equiv \sum_{i,j} Y_i Y_j \lambda_{ij}(s,T) - Y_sY_i \lambda_{si}(k,T),
\label{rate_eqn}
\end{equation}
where $\lambda_{ij}(s,T)$ is the rate at which species $i$ and $j$ react to form species $s$ at a temperature $T$.

All reaction rates were taken from Glover \& Abel (2008). Because of the intrinsic order of magnitude spread in these rates, the resulting equations are `stiff,'  meaning  that the ratio of the minimum and maximum eigenvalue of the Jacobian matrix, $ J_{i,j} = \partial \dot Y_i / \partial Y_j$, is large and imaginary. This means that implicit or semi-implicit methods are necessary to efficiently follow their evolution.  To address this problem, we arranged the molar fractions of the 6 species (excluding e$^-$) into a vector and solved the resulting system of equations using a $4^{th}$ order Kaps-Rentrop, or Rosenbrock method
(Kaps \& Rentrop 1979) as described in more detail in Gray \& Scannapieco (2010).

To ensure the stability of the chemistry routine  while at the same time allowing the simulation to proceed at the hydrodynamic time-step,  we developed a method of cycling over multiple Kaps-Rentrop time steps within a single hydrodynamic time step. To do this we estimated an initial chemical time step of each species as 
\begin{equation}
	\tau_{{\rm chem},s} = \alpha_{\rm chem} \frac{{Y}_s+0.1Y_{H^+}}{\dot{Y_s}}, 
\label{tau_chem}
\end{equation}
where $\alpha_{\rm chem}=0.5$ is a constant that controls the desired fractional change of the fastest evolving species. Note that this estimate is offset by adding a small fraction of the ionized hydrogen abundance to deal with conditions in which  a species is very low in abundance but changing rapidly. 
	
Once calculated, these species are compared to each other and the $\tau_{{\rm chem},s}$ associated with the fastest evolving species is chosen as the subcycle time step, $\Delta t_{\rm sub}.$ If this is longer than the hydrodynamic time step, the hydrodynamic time step is used instead and no additional subcycling is carried out. If subcycling is required, the species time step is subtracted from the total hydrodynamic time step and the network is updated over the chemical time step.  At this point we implement
the change in internal energy due to ionizations and recombinations, which is given by 
\be
\Delta t_{\rm sub} \dot E_{\rm chem} = - N_A \sum E_s \Delta Y_s,
\ee
where $E_s$  and $\Delta Y_s$ are the binding energy and change in the molar fraction of species $s$ over $\Delta t_{\rm sub}$,
and $N_A=6.022\times10^{23}.$   Then the cooling routine is called and given the new internal energy the temperature is updated, and
the species time steps are recalculated and compared to the remaining hydrodynamic time step. Given a new $\Delta t_{\rm sub},$
the cycle is repeated until a full hydrodynamic time step is completed.  Finally, in cases in which the gas is $\ge 10^5$K  H and He are always fully ionized, avoiding the need for matrix inversions. 

\section{Results}

\subsection{Scaling Relations and Energy Balance}

The timescale of energy to decay in a turbulent cascade is given by $t_{\rm diss} \approx L/V,$ where $L$ and $V$ are the length scale and velocity of the largest turbulent eddies, which in our simulations are $\approx L_f$ and $\sigma_{\rm  3D,M} = 3^{1/2} \sigma_{\rm  1D,M}$ respectively.  This means that if we were to increase the physical scale of the entire simulation by a factor $X,$ leaving everything else the same, the timescale for turbulent energy decay would also be increased by a similar factor $X$.

In a steady state configuration, this energy input is balanced by the radiative cooling of the gas, which is  proportional to the mean gas density, $\bar n.$  This means that if we take a system in equilibrium and rescale the  domain by a factor $X$ while at the same time decreasing the mean density by a factor $X$, the overall timescale will increase but the system will otherwise remain the same.   For this reason, rather than tie our simulations to a fixed length scale, we report our results in units of $H,$  and dynamical times $t_{\rm dyn} \equiv H/\bar \sigma_{\rm 3D,M}$, where $\bar \sigma_{3D,M}$ is the time averaged mass-weighted three-dimensional velocity dispersion.

Furthermore, as the decay rate of the turbulence is proportional to the driving scale rather than the scale height, and our simulations are unable to achieve values of $H/L_f$ as large as encountered in real galaxies, for any equilibrium configuration the mean gas densities in our simulations will be lower than those in a real galaxy.   On the other hand, as discussed in more detail below, the behavior of the interstellar medium is most likely determined by the range of temperatures achieved in the simulation, which in turn is determined by the mean velocity dispersion.   Thus, while we report $H \bar n$ in each of our simulations, we expect $\bar \sigma_{\rm 1D, M}$ to be the key observational parameter used to compare them against the behavior of real galaxies, and so we name each run according to this value.  A summary of $\bar \sigma_{\rm 1D,M},$ $H \bar n,$ and the other important parameters from each of our runs is given in Table 1.

\begin{figure}
\epsscale{1.2}
\plotone{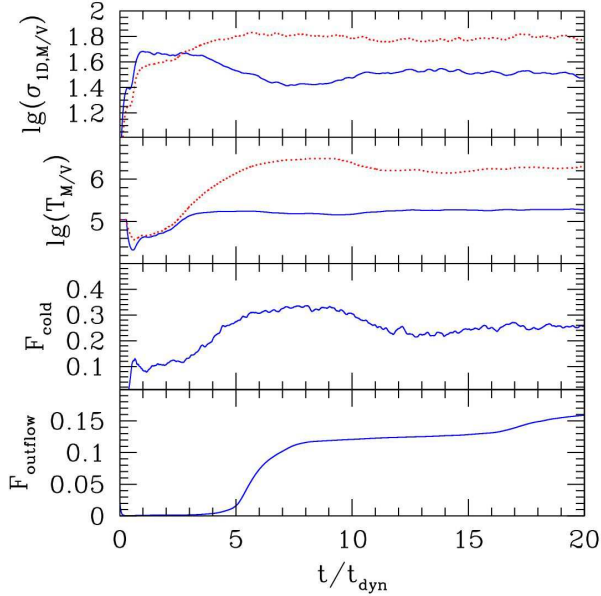}  
\caption{Evolution of global quantities in our fiducial run, S34.   {\em Top Panel:} Mass averaged velocity dispersion (solid line) and volume averaged velocity dispersion (dotted line) in units of km/s.  {\em Second Panel:} Mass averaged temperature (sold line) and volume averaged temperature (dotted line) in units of K. {\em Third Panel:} Mass fraction of cold ($T \leq 10^{4.1}$K) gas. {\em Bottom Panel:}
Mass fraction of gas leaving the simulation volume.  In all panels time is expressed in units of $t_{\rm dyn} = 3^{1/2} H/\bar \sigma_{\rm 1D,M}$ where $\bar \sigma_{\rm 1D,M}$ = 34 km/s.}
\label{fig:evolution01}
\end{figure}

\begin{table*}[!ht]
\caption{Run Parameters}
\vspace*{0.2cm}
\begin{center}
\begin{tabular}{c | c c c c c c c c }
\hline\hline
Name  & $\bar \sigma_{1D,M} $ & $\bar \sigma_{3D,M}$  & $H \bar n^{\rm(a)}$ & $t_{\rm dyn} \bar n$   & $T_{\rm grav}$ & $H/\Delta x$ &  $t_{\rm final}$  & $F_{\rm outflow}$ \\
   & (km/s) & (km/s) & $ (M_\odot \, \rm pc^{-2}) $ & (yr cm$^{-3}$) & K & &   ($t_{\rm dyn}$)  & \\
\hline
S34     &  34 	& 59 	                  & 	0.061   & 	$7.5 \times 10^4$  & $1 \times 10^5$ & 64 & 	20 & 0.16  \\ 
S20       &  20 	& 35	          & 	0.0061 & 	$1.3 \times 10^4$ & $1 \times 10^5$ & 64 & 	40 & 0.02  \\  
S29       & 29      & 50 	          & 	0.018   & 	$2.6 \times 10^4$ &  $1 \times 10^5$ & 64  & 	20 & 0.01  \\  
S61       & 61     & 106 	& 	0.18     & 	$1.2 \times 10^5$ & $1 \times 10^5$ & 64 &	10 & 0.95  \\ 
S59G    & 	59    & 102	&          0.18   &  $1.3 \times 10^5$   &  $3 \times 10^5$ & 64 &      10 & 0.05   \\  
S35HR & 35    & 60	          & 	0.061   & 	$7.3  \times 10^4$  & $1 \times 10^5$ & 96 &    12 & 0.27  \\  
\hline
\end{tabular}
\end{center} \par
{\footnotesize $^{\rmÊ(a)}$ Note that the column depth at a given velocity dispersion will be significantly less than in a real galaxy due to our inability to adopt a turbulent driving scale smaller than $L_f/H = 0.1.$}

\label{table:grid}

\end{table*}

\subsection{Fiducial Case}

\begin{figure*}
\epsscale{1.2}
\plotone{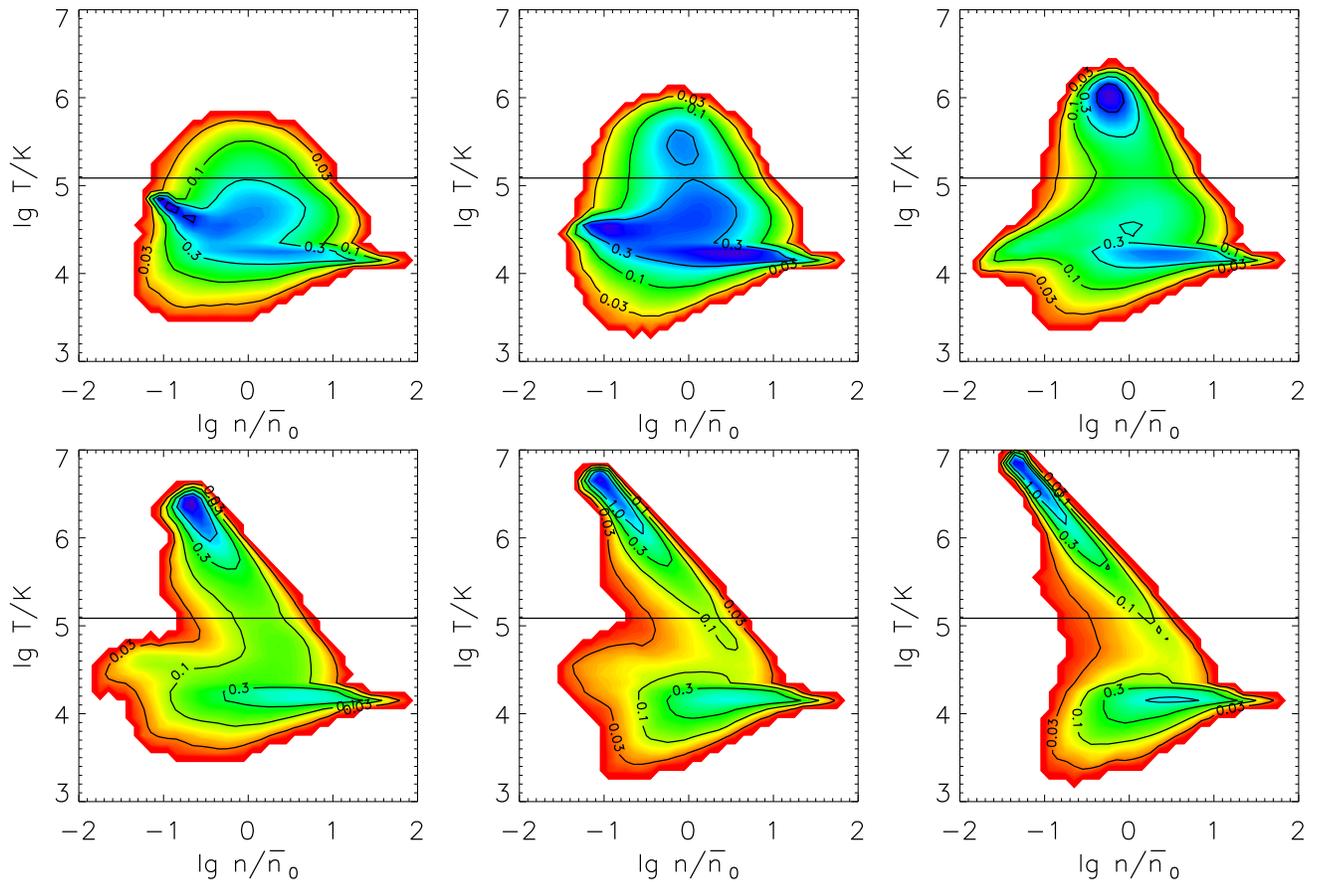}  
\caption{Phase diagram showing logarithmic contours of the mass-weighted PDF of the gas distribution in run S34 at times $t/t_{\rm dyn}=$ 0.98, 2.0, and $3.0$ (top row) and $t/t_{\rm dyn}=$ 4.0, 5.0, and 6.0 (bottom row).  In each panel the density is in units of the initial mean density, the temperature is in units of K, and the horizontal line corresponds to $T_{\bar \sigma_M}$ as given by eq.\ (\ref{eq:temp}) with $\bar \sigma_{\rm 1D,M}$ = 34 km/s.  All contours are labeled by their values relative to the PDF bin with the most mass.}
\label{fig:pdfd01}
\end{figure*}

\begin{figure*}
\epsscale{1.1}
\plotone{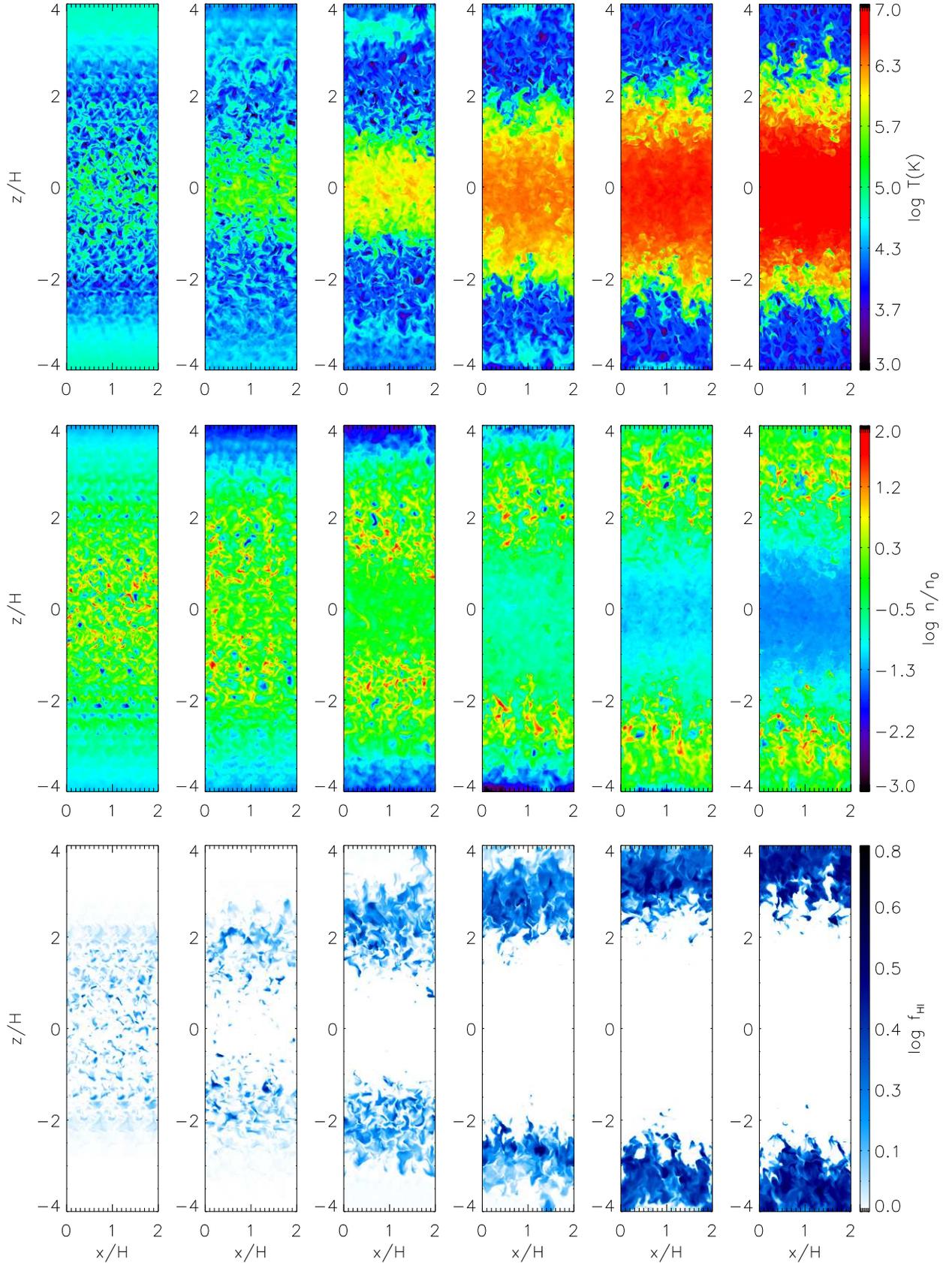}  
\caption{Log of the temperature distribution in K (top), the log of the  density distribution in units of the initial mean density (center), and the neutral hydrogen fraction (bottom) in a vertical slice through run S34 at times corresponding to those in Fig.\ \ref{fig:pdfd01}.  From left to right $t/t_{\rm dyn}=$ 0.98, 2.0, 3.0, 4.0, 5.0, and 6.0, and lengths are labeled in units of $H.$}
\label{fig:tdh01}
\end{figure*}

In Fig.\ \ref{fig:evolution01} we show the evolution of the global mean quantities in our fiducial run, S34.  After the short initial stirring stage, cooling is turned on and the temperature begins to drop significantly, leading to $F_{\rm cold} \approx 0.1$ and an increase in $\sigma_{1D,M}$ to $\approx$ 45 km/s.  Equilibrium is then maintained for roughly three dynamical times, over which $\sigma_{\rm 1D,M}$ and $\sigma_{\rm 1D,V}$ are roughly equal and constant.  During this time, the mass and volume averaged temperatures $T_M$ and $T_V$ are similar, and approximately equal to the postshock temperature for a $v=\bar \sigma_{\rm 3D,M}$ shock
\be
T_{\rm \bar \sigma_M} = {\mu m_p\over 2 k_B} \bar \sigma_{\rm 3D,M}^2 = 100 {\rm K} \, \frac{\bar \sigma_{\rm 1D,M}^2}{\rm km^2 \, s^{-2}},
\label{eq:temp}
\ee
where $k_B$ is Boltzmann's constant and $\mu=0.6$ is the molecular weight of the gas.

This quasi-stable phase is fleeting however, as after approximately three dynamical times, the volume averaged temperature begins to climb dramatically, increasing by over an order of magnitude to $3 \times 10^6 K$, by $t/t_{\rm dyn} \approx 6.$   Furthermore, this rapid proliferation of hot gas is accompanied by a rapid ejection of material, and over 10\% of the gas mass is driven out from the simulation volume from $t/t_{\rm dyn}$ = 3 to 6.

To explore the cause of this dramatic heating, in Fig.\ \ref{fig:pdfd01} we present gas phase diagrams for times $t/t_{\rm dyn} \approx 1$ to $6$. To construct these distributions, we partitioned the gas  into equally-spaced logarithmic bins in temperature and density, and computed the total mass contained in each bin.   Initially, the majority of the gas is at or below, $T_{\rm \bar \sigma_M},$ the postshock temperature corresponding to the mass averaged velocity dispersion, and only a small fraction of the gas is at somewhat higher temperatures.   As time progresses, however, a secondary peak develops in the PDF, which appears first at $\approx 3 \times 10^5K$ and then moves to higher temperatures and lower densities as time progresses, traveling along a constant pressure trajectory with $T \propto n^{-1}.$   

The key to understanding this behavior is to compare the local cooling time of the gas to the dynamical time.   To facilitate this comparison, we first write the dynamical time in the same units as the cooling times shown in Fig.\ 1, that is
\be
t_{\rm dyn} \bar n = \frac{3^{1/2} H \bar n}{\bar \sigma_{\rm 1D,M}} = 1.2 \times 10^6 \, {\rm yr} \, {{\rm cm}^{-3}} \frac{H \bar n}{M_\odot \, {\rm  pc}^2}\frac{35 \, {\rm km \, s^{-1}}}{\bar \sigma_{\rm 1D,M}},
\label{eq:tdyn}
\ee
which for run S34 gives $t_{\rm dyn} \bar n = 7.2 \times 10^4  \, {\rm yr}  \, {\rm cm}^{-3}.$   This is roughly 10 times the cooling time of $10^5$K gas, which is to be expected as, in equillibrium, gas cooling balances turbulent energy input, and the dissipation timescale for turbulence is smaller than the dynamical time by a factor of $H/L=10.$ 

However, as it is a stochasitic process, turbulent driving does not heat the gas uniformly, but rather deposits only a small amount of energy in some regions, while heating other regions to several times the mean temperature, an effect is stronger in supersonic flows, such as those in the ISM (Pan \&  Padoan 2009, Pan et al.\ 2009, Pan \& Scannapieco 2011).  As $t_{\rm cool}$ rises sharply above $\approx 2 \times 10^5$K, the portion of the gas that is heated to above this value takes several turbulent dissipation timescales to cool.  This means that before this overheated gas can settle back down to the mean temperature, subsequent turbulent energy input will heat it even further, causing it to expand to maintain pressure equilibrium with its surroundings.   Both the increased temperature and lower density lead to even longer cooling times, which rapidly become orders of magnitude larger than the turbulent driving time, and even the dynamical time.  The result is the rapid, runaway heating of the hot, low-density gas, which pushes its way out of the galaxy, dragging a substantial portion of the total gas mass with it.  In other words, {\em the ejection of material from the disk is a direct consequence of turbulent support at the level assumed in this simulation.}

Note however, that whether this ejected gas is able to escape from the host galaxy will depend on the overall gravitational potential.  In the case of our fiducial simulation $v_{\rm esc} = \sqrt{2 g_0 4 H}$ = 108 km/s which is comparable to the vertical thermal motions of $10^6 K$ ionized gas, resulting in a substantial mass of outflowing gas.    Note that, assuming a Gaussian distribution of random velocities, since $v_{\rm esc}/\sigma_{1D} = 3.2,$ less than 0.2\% of the gas would be accelerated outward above escape velocity 
in the absence of runaway heating.  On the other hand, a galaxy with a higher escape velocity, corresponding to either a larger vertical simulation volume or a higher value of $T_{\rm grav}$, and would drive material out of the midplane without unbinding it from the host galaxy, as studied in more detail below.  

Note also  that at the same time the high-temperature runaway  occurs in this simulation, the cold fraction shown in Fig.\ \ref{fig:evolution01} increases appreciably.  This can be understood in terms of a simultaneous runaway occurring in the cold, dense gas.  For this gas, the typical turbulent velocity is well above the local sound speed, meaning that it is susceptible to shocks with relatively high Mach numbers.  Furthermore, because of the short cooling times in these regions, these will be strong, radiative shocks that lead to substantial density enhancements. Thus in Fig.\ \ref{fig:pdfd01}, we see a pile up of cold material at 10-100 times the mean density, even as the hottest gas moves to lower densities and ever-increasing temperatures.  This means turbulent runaway is a two-way process in which inefficient cooing in the hottest regions is accompanied by compression-driven cooling in the coldest regions.

\begin{figure}
\epsscale{1.2}
\plotone{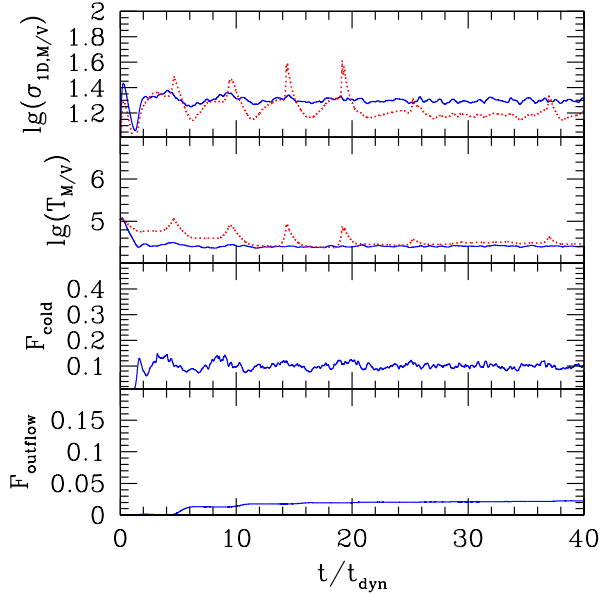}  
\caption{Evolution of global quantities in a run with a low turbulent velocity dispersion, S20.  Panels and lines are as in Fig.\ \ref{fig:evolution01}, and time is expressed in units of $t_{\rm dyn} = 3^{1/2} H/\bar \sigma_{\rm 1D,M},$ where $\bar \sigma_{\rm 1D,M}$ = 20 km/s.}
\label{fig:evolution001}
\end{figure}

 \begin{figure*}
\epsscale{1.2}
\plotone{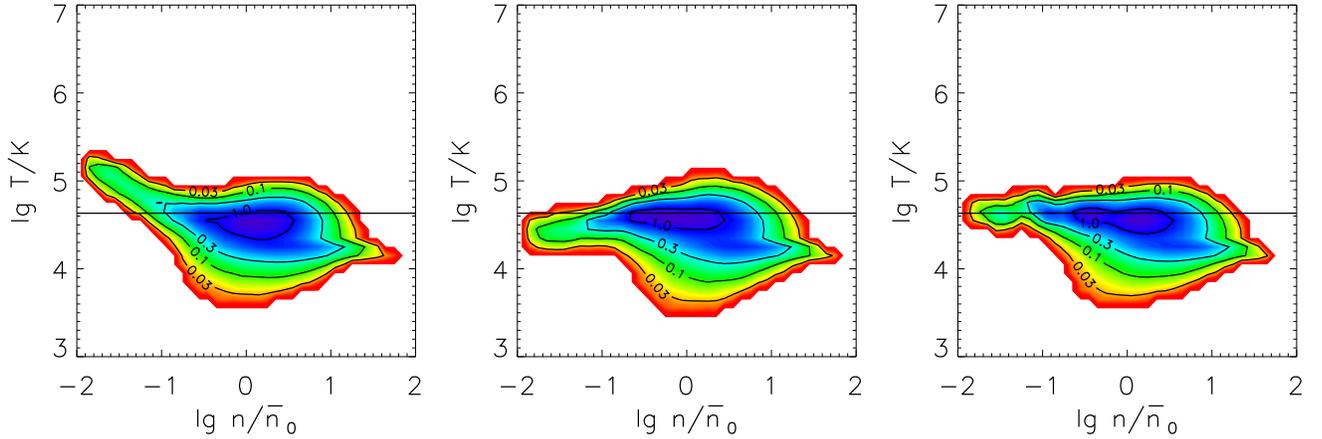}  
\caption{Phase diagram showing logarithmic contours of the mass-weighted PDF  of the gas distribution in run S20 at times $t/t_{\rm dyn}=$ 10,  20, and 30.  As in Figure \ref{fig:pdfd01},  the density is in units of the initial mean density, the temperature is in units of Kelvin, and the horizontal line  corresponds to $T_{\bar \sigma_M,}$ for $\sigma_{\rm 1D,M}$= 20 km/s as given by eq.\ (\ref{eq:temp}).}
\label{fig:pdfd001}
\end{figure*}

\begin{figure*}
\epsscale{1.2}
\plotone{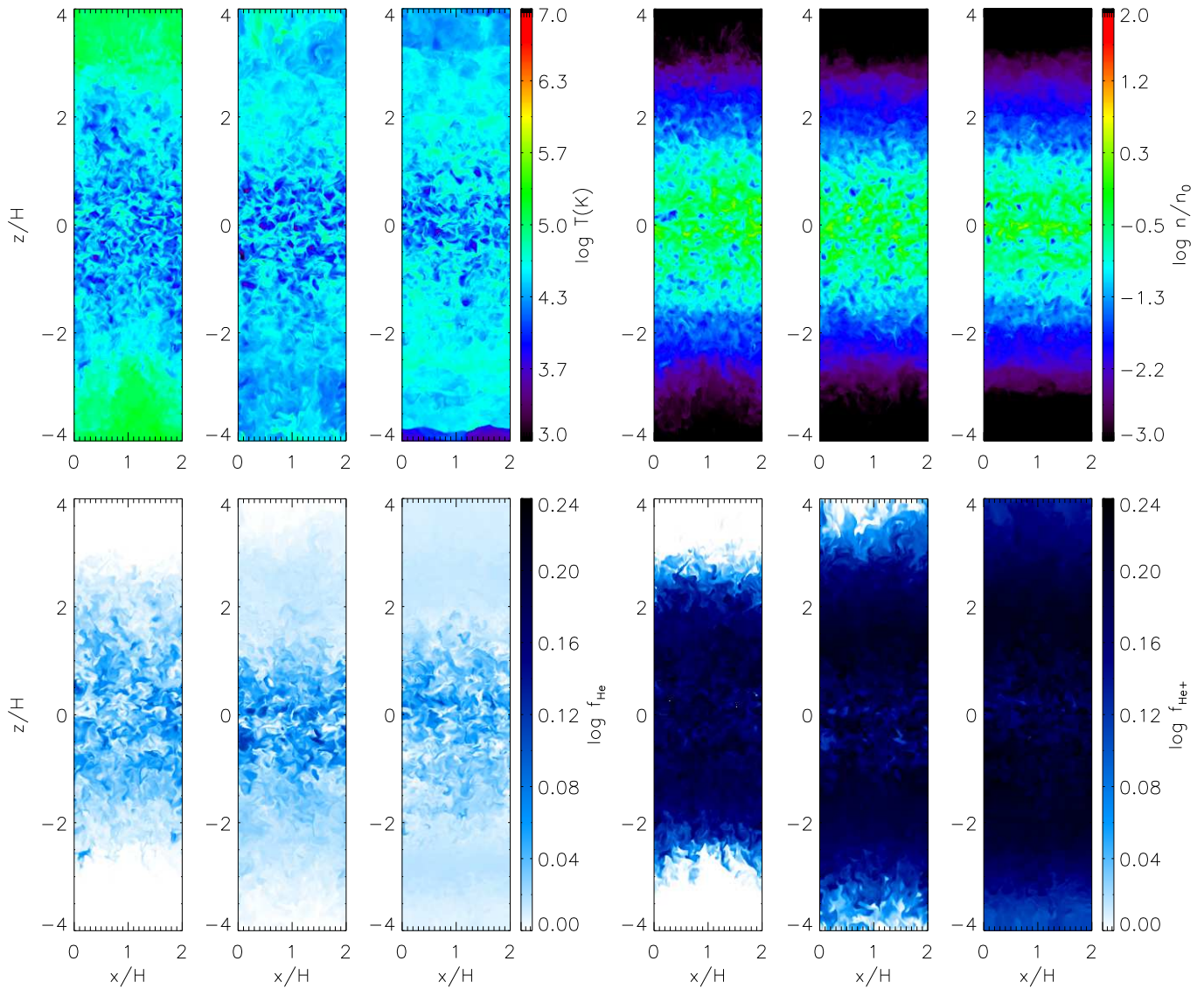}  
\caption{Log of the temperature (top left panels) and density (top right panels) distributions, and the fraction of neutral He (bottom left panels) and He$^+$ (bottom right panels) in a vertical slice from run S20 at times $t/t_{\rm dyn}=$ 10,  20, and 30.}
\label{fig:tdhe001}
\end{figure*}

The development of this turbulent runaway  is further illustrated in Fig.\  \ref{fig:tdh01}, which shows the temperature, density, and neutral hydrogen mass fraction in a vertical slice during the same times as the phase diagrams in Fig.\ \ref{fig:pdfd01}. At $t/t_{\rm dyn} = 1$, temperature contrasts are modest with all the gas well below $10^6$K throughout the galaxy.  Likewise, the density distribution is similar to the initial hydrostatic configuration, and dense clumps of gas are seen only near the midplane, accompanied by regions of neutral hydrogen.   By $t/t_{\rm dyn} = 2$, isolated patches with temperatures approaching $10^6$K are visible, and hydrogen near the midplane has been dissociated, but the overall temperature and density distributions are largely unchanged. 

By $t/t_{\rm dyn} = 3$, on the other hand, a dramatic change has begun.   Patches of $T \ge 2 \times 10^6$K are visible near the midplane, increasing the pressure to the point that the dense clumps of gas are rapidly moved outward, and neutral hydrogen is found only at high altitudes.    From $t=4$ onward, heating progresses catastrophically, increasing the temperature of the gas near the midplane up to $\approx 10^7$K, and driving dense clumps of gas to the edges of the simulation and beyond.   

\subsection{Low-energy Case}

 In contrast to our fiducial run, Fig.\ \ref{fig:evolution001} illustrates the evolution of the global quantities in a run in which the density, and consequently the cooling and turbulent driving rates, are reduced by a factor of ten.   In this case, the average mass-weighted velocity dispersion is only $\bar \sigma_{\rm 1D,M} = 20$ km/s and the corresponding post-shock temperature is $T_{\bar \sigma_M} = 4.2 \times 10^4$K.  After an initial cooling phase, the mass averaged temperature settles down to approximately this value, maintaining it for the 40 dynamical time we simulated, twice as long as our full fiducial simulation.   Likewise, while some fluctuations are seen in the volume averaged temperature and velocity dispersion, which are more sensitive to the low density regions near the edge of the simulation, even these settle down almost completely after 20 dynamical times.    The cold fraction remains almost constant at 10\% from about 10 dynamical times onwards, and only 2\% of the mass leaves the simulation volume in the initial rearrangement of the medium.    Unlike the fiducial case, the medium has settled into a stable equilibrium that can be maintained indefinitely.
 
Again, this behavior can be directly understood from the mass weighed PDF, shown in Fig.\  \ref{fig:pdfd001}.  Unlike the fiducial case, the gas phase distributions shown in this plot falls entirely below the $2 \times 10^5$ limit above which $t_{\rm cool}$ rises rapidly.    Since the cooling time is roughly constant below $2 \times 10^5$, the cooling time of overheated gas is comparable to the average cooling time, which in turn is comparable to the turbulent driving time.   This means that the gas will have  sufficient time to cool back down to the mean temperature before further turbulent energy input takes place.  Furthermore, as turbulent velocities are significantly smaller than in the fiducial run, the shocks in the cold regions are much weaker, leading to small density enhancements that fail to cause a substantial increase in $F_{\rm cool}$.

In addition, the chemical reactions included in our simulation further stabilize the mixture of neutral and ionized gas that occurs in this temperature range.   This is because when cold gas is heated it dissociates, removing internal energy from the system and reducing the increase in the temperature.   Likewise when hot gas is cooled it recombines, adding internal energy to the system that counteracts the cooling.

Profiles of the temperature, density, and He and He$^+$ mass fractions in a vertical slice through this simulation are shown in Fig.\ \ref{fig:tdhe001}.      The temperature fluctuations are minor and dense gas is strongly concentrated near the midplane at all times, such that it is extremely convectively stable. At all altitudes, the gas is a mixture of neutral and ionized material, and the majority of helium is singly-ionized.  This singly-ionized helium will recombine and release internal energy if it is cooled and dissociate and extract internal energy if   it is heated.  In contrast to the run with more vigorous turbulence, the galaxy has found a  profile that is stabilized from significant changes over the lifetime of the system.Ê

\subsection{Intermediate and High-Energy Cases}

\begin{figure*}
\epsscale{1.15}
\plotone{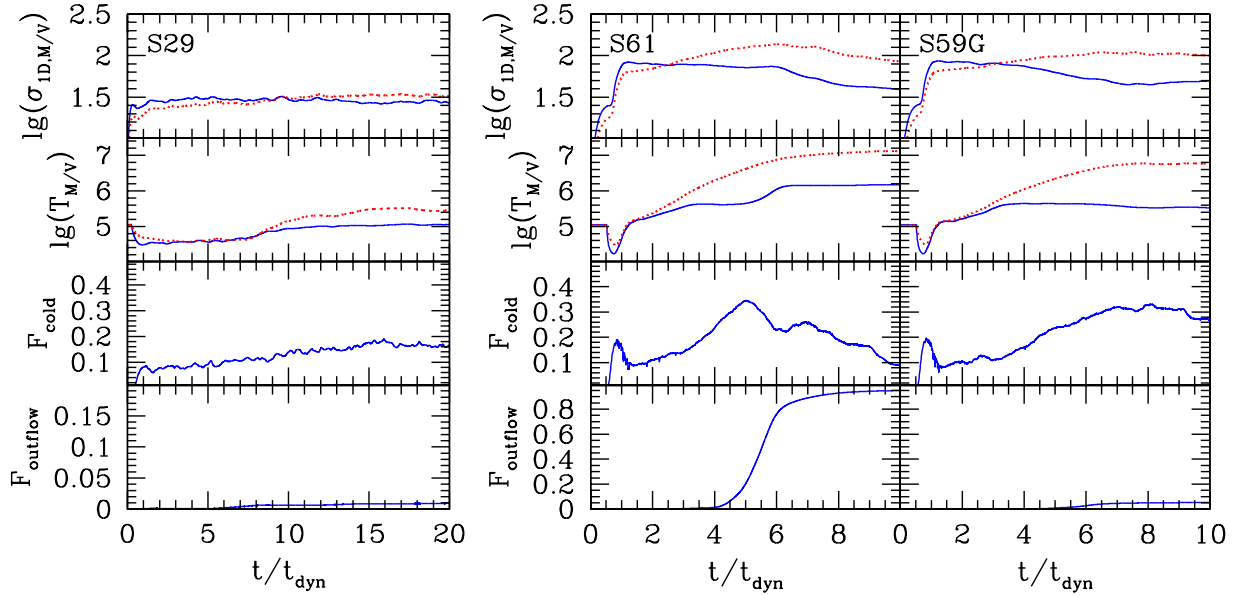}  
\caption{Evolution of global quantities in a run with intermediate turbulent velocity dispersion, S29 (left column), a run with very large velocity dispersion, S61 (center panels), and a run with very large velocity dispersion and increased gravity, S59G (right column).     In each column, panels and lines are as in Figs.\ \ref{fig:evolution01} and \ref{fig:evolution001}, and time is expressed in units of $t_{\rm dyn} = 3^{1/2} H/\bar \sigma_{\rm 1D,M}$ where, from left to right, $\bar \sigma_{\rm 1D,M}$ = 29 km/s, 61 km/s, and 59 km/s respectively.}
\label{fig:evolution_0.03_0.3}
\end{figure*}

Next we consider an intermediate model, S29, in which the  density is 30\% of  that in the fiducial run  but 3 times that in the low-energy run. In this case, whose global quantities are plotted in the left column of Fig.\ \ref{fig:evolution_0.03_0.3}, the mass and volume averaged velocity dispersions quickly settle into a roughly constant configuration, which is maintained throughout the simulation.  Here $\bar \sigma_{\rm 1D,M} = 29$ km/s, and while there is some indication of a shift in the volume averaged temperature to higher values, this increase is much less than seen in run S34.  Furthermore the cold fraction is relatively stable and there is very little material lost from the simulation volume.

\begin{figure*}
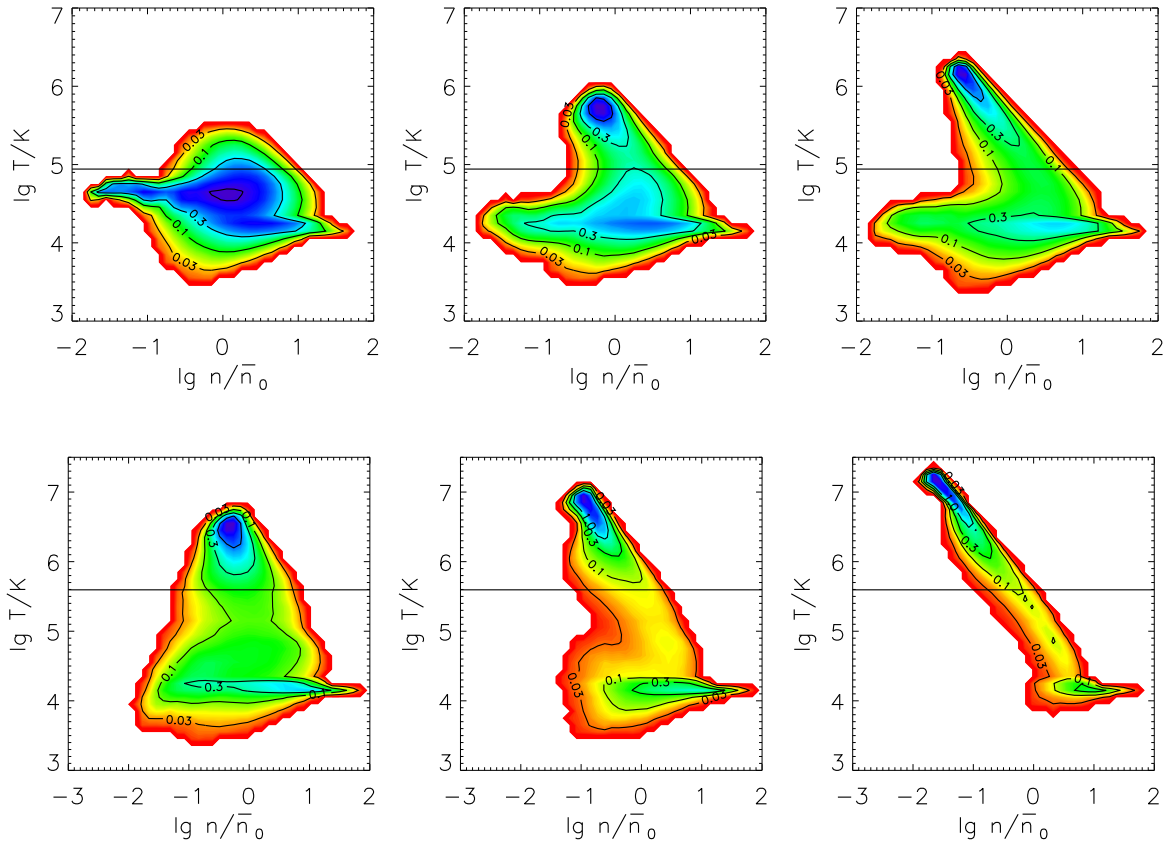

\epsscale{1.1}
\plotone{f9a.eps}  
\plotone{f9b.eps}  
\caption{{\em Top:} Phase diagram showing logarithmic contours of the mass-weighted PDF  of the gas distribution in intermediate-energy run S29 at times $t/t_{\rm dyn}=$ 3.0,  9.1, and 15, as compared to $T_{\bar \sigma_M}$ with $\bar \sigma_{\rm 1D,M}=29$ km/s.  {\em Bottom:} Phase diagram showing logarithmic contours of the mass-weighted PDF of the gas distribution in the very high-energy run S61 at times $t/t_{\rm dyn}=$ 2.9, 4.5, and 6.0, as compared to $T_{\bar \sigma_M}$ with $\bar \sigma_{\rm 1D,M}=61$ km/s.  Note that to accommodate the more extreme conditions in this run the maximum temperature in this row $3 \times 10^7$K and the minimum value of $\lg(n/\bar n_0)$ is $-3.$}
\label{fig:pdfd_0.03_0.3}
\end{figure*}

An inspection of the mass-weighed PDF, shown in the top row of Fig.\  \ref{fig:pdfd_0.03_0.3},  indicates that while a significant fraction of the gas is driven to $10^6$K, where the cooling time is comparable to the dynamical time of $2.5 \times 10^4$ yr cm$^{-3}$, this overheated gas falls just short of achieving the turbulent runaway seen in the fiducial run.  All in all, this run appears to be at the boundary of the transition between stable and unstable turbulent support.  Thus it is likely that  $\bar \sigma_{\rm 1D,M}  \approx 35$ km/s represents a critical value above which turbulent support will become unstable.

A very high energy simulation,  in which the density is increased by a factor of 3 from the fiducial run, confirms that $\bar \sigma_{\rm 1D,M}  \approx 35$ km/s is a critical threshold.    The global evolution from this run, S61, is plotted in the center panel of Fig.\ \ref{fig:evolution_0.03_0.3}, which shows an unstable evolution even more extreme than in the fiducial run.   After only one dynamical time, the volume averaged temperature begins to exceed the mass averaged temperature, and within three dynamical times it moves into the highly unstable $\geq 10^6$K range.   At this point, as in the fiducial run,  the combination of  long cooing times and the rapid energy input from turbulent decay cause the gas to quickly be heated to even higher temperatures, in this case exceeding $10^7$K.   The result is an extremely rapid expansion of the medium, and by $t/t_{\rm dyn}=6$ over 90\% of the gas flows out of  the simulation volume.

Again, each of these phases is clearly visible in the mass-weighted PDF, shown in the bottom row of Fig.\  \ref{fig:pdfd_0.03_0.3}.  Here we see that at $t/t_{\rm dyn}=2.9$ a large fraction of the gas has collected at $\approx 2 \times 10^6$K, well above the average post-shock temperature corresponding to the velocity dispersion.   By  $t/t_{\rm dyn}= 4.5$ the temperature of the overheated material increases by  an order of magnitude,  moving along a $n \propto T^{-1}$ line of constant pressure.  Finally, by $t/t_{\rm dyn}= 6.0$ a large fraction of the gas has flowed out of the simulation volume, moving the histogram to the left.

\begin{figure}
\epsscale{1.2}
\plotone{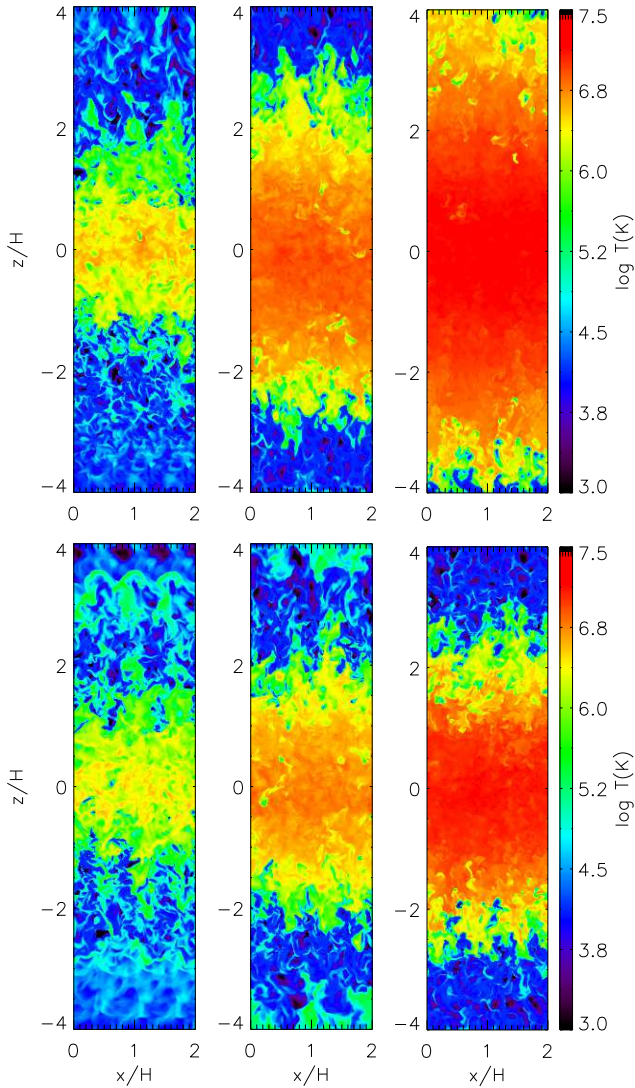}  
\caption{Log of the temperature distribution in run S61 (top panels) at times $t/t_{\rm dyn} =$ 2.9, 4.5, and 6.0, as contrasted with those from a similar run with higher gravity, S59G (bottom panels) at times $t/t_{\rm dyn} = $ 3.0, 4.4, and 6.0.  Note that the upper temperature in these plots is $3 \times 10^7$K, rather than $10^7$K as used in plotting lower-energy runs in Figs.\  \ref{fig:tdh01}, \ref{fig:tdhe001}, and \ref{fig:evolution_0.03_0.3}.}
\label{fig:tcomp}
\end{figure}

Note that even without turbulent runaway, one would have expected this gas to escape from the gravitational potential, as the turbulent velocity is high enough that the average mass-weighted temperature quickly rises to $T_{\rm} = 3 \times 10^5$K, while the original gravitational potential was established to be in hydrostatic equilibrium with $T = 10^5$K.  Thus we carried out a second, very high energy run, S59G, in which all parameters were the same as in run S61, but the gravity was increased by a factor of 3 such that $T_{\rm grav} = 3 \times 10^5$K, and $v_{\rm esc}  = 187$ km/s.

In this case, for which the global quantities are plotted in the rightmost panels of Fig.\ 8, the evolution of the mass averaged and volume averaged velocity dispersions, and mass averaged and volume averaged temperatures is extremely similar to run S61.  In particular,  the volume averaged temperature again begins to exceed the mass averaged temperature within a dynamical time, again within three dynamical times it moves into the highly unstable $\geq 10^6$K regime, and again it finally moves off to higher temperatures approaching $10^7$K.  However in this run, this runaway is not able to drive material off of the simulation grid, and the escape fraction after 10 dynamical times is only 5\%.

 Fig.\ \ref{fig:tcomp} contrasts the temperature evolution in a vertical slice from this run with the evolution of a similar slice from the lower gravity run, S61.  Note that outputs at slightly different $t/t_{\rm dyn}$ are compared in this figure  because the dynamical time is dependent on the average velocity dispersion, which can not be  precisely predicted before each run. In both cases, the initially slow build-up of overheated gas near the galaxy midplane leads to a rapid runaway that pushes dense gas several scale-heights outward.  Only in the lower gravity case, however, is this runaway sufficiently powerful to drive the gas out to distances greater than $4H,$ while in the high gravity case, it reaches heights of only $\approx 2H$ above the disk, where it remains until we stopped the simulation at 10 dynamical times.   This means that while the ejection of gas from the disk is  a direct consequence of the level of turbulent support, whether or not this gas will escape from the galaxy is dependent on the strength of the gravitational potential.   In a small galaxy with high $\bar \sigma_M,$ the gas would escape into the intergalactic medium in a powerful galaxy outflow.  In a large galaxy with high $\bar \sigma_M$, on the other hand, one would expect the presence of a ``galactic fountain" type circulation flow in which the gas flows upwards from regions with vigorous turbulence like the one were have simulated, but then moves laterally at high altitude and falls back onto the disk in less turbulent regions  (\eg Shapiro \& Field 1976; Bregman 1980; Houck \& Bregman 1990; Walters \& Cox 2001),

\begin{figure}
\epsscale{1.2}
\plotone{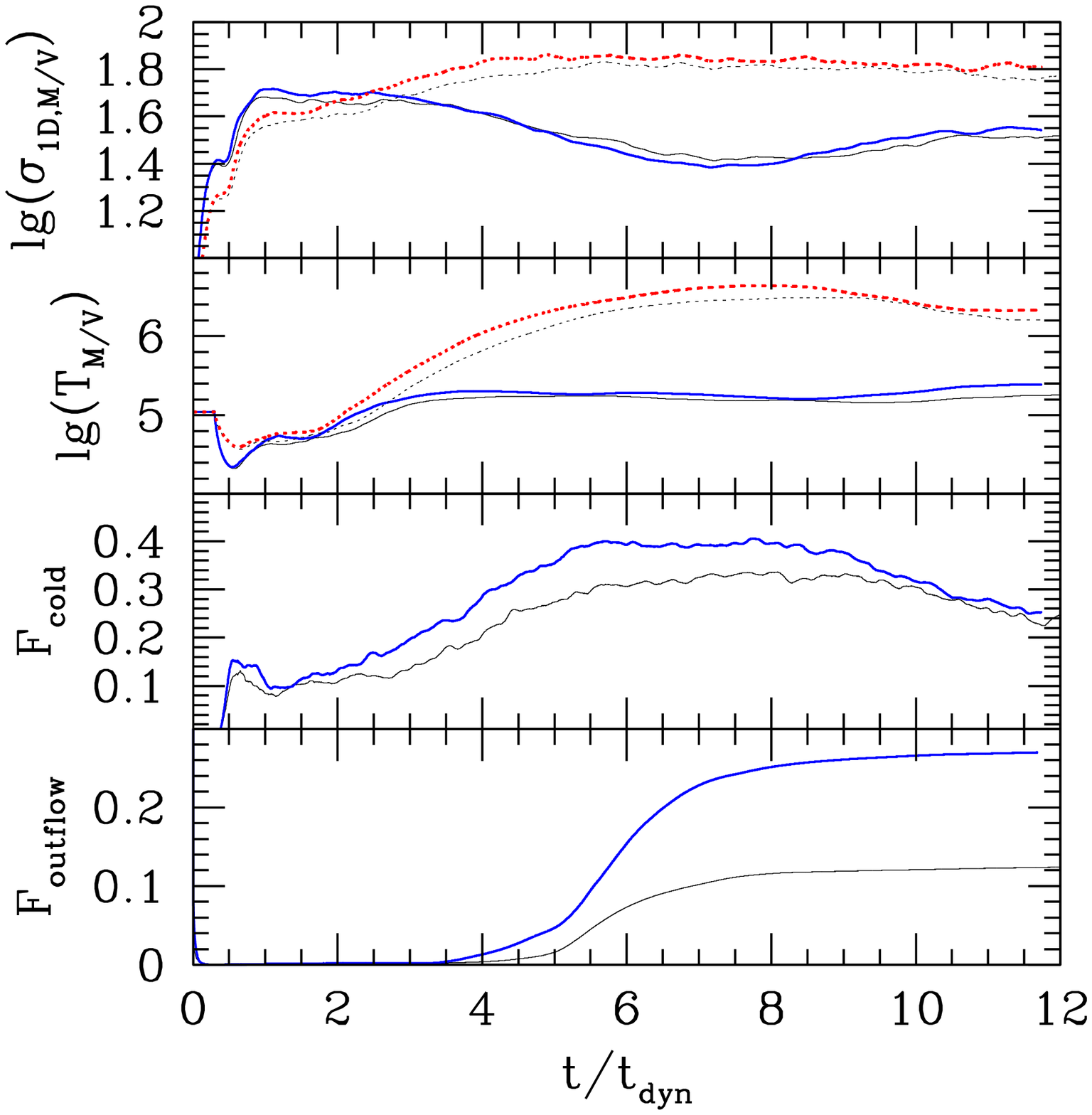}  
\caption{Evolution of global quantities in run S35HR, a high resolution version of the fiducial run.  Panels and lines are as in Figs.\ \ref{fig:evolution01} and \ref{fig:evolution001}, and time is expressed in units of $t_{\rm dyn} = 3^{1/2} H/\bar \sigma_{\rm 1D,M}$ where $\bar \sigma_{\rm 1D,M}$ = 35 km/s.
In each panel the thick (colored) lines are from run S35HR, while the thin black lines are for comparison and are taken from the fiducial run, S34.}
\label{fig:evolution_res}
\end{figure}

\subsection{Resolution}

Finally, we assess the impact of resolution on our results.  As $L_f/\Delta x = 6.4$ in the simulations described above,  a lower-resolution comparison test was not possible because the turbulent driving scale would be insufficiently well resolved.  However we carried out a higher resolution simulation in which all parameters were matched to the fiducial (S34) run, but $H/\Delta x$ was increased from 64 to 96.  In Figure \ref{fig:evolution_res}, the evolution of the global quantities from this run, S35HR, are contrasted with those from the fiducial case.    Here we see that the evolution of both temperature and velocity dispersion is very similar in both runs, generating a turbulent runaway that commences at almost exactly the same time and reaches almost exactly the same final temperature. 

On the other hand, the fraction of the gas flowing out of the simulation volume is significantly higher in this run.   On closer inspection of the gas PDF (which is not shown here) this appears to be due to the fact that the higher resolution run, which is less susceptible to numerical diffusion, is able to span a slightly larger range of temperatures than the fiducial run.   While this difference has only a minor effect on the average quantities, it does lead to a noticeable increase both in the fraction of cold, $T \leq 1.5 \times 10^4$K gas, and the fraction of $T \geq 2 \times 10^5$ gas that participates in the turbulent runaway.    As the numerical resolution increases, there are two effects that are likely to broaden the temperature distribution in this manner. The first is that numerical diffusion becomes  weaker with increasing resolution, reducing the mixing between hot and cold regions.   The second is that energy dissipation in turbulent  flows is intermittent, meaning that the range of dissipation rates broadens at smaller physical scales, causing the heating rate to have a broader PDF as resolution increases  (Pan \&  Padoan 2009, Pan et al.\ 2009) 

Together these effects lead to a broader temperature PDF in run S35HR, leading to a larger amount  of gas at extreme temperatures, both hot and cold. The increase in hot gas fraction, in particular, is enough to expand the overall gas distribution somewhat more than in the fiducial run, moving a larger overall fraction of the gas off of the grid. Thus we conclude that while the temperature and density structure of the turbulent runaway are well resolved in our simulations,  the final vertical distribution is weakly dependent on resolution and strongly dependent on the  assumed gravitational profile.

\section{Discussion}

Taken together, our simulations imply that while there is more to be learned about the details of gas escape from the overall gravitational potential, we can nevertheless be confident in our identification of a fundamental transition in stratified,  turbulently-stirred, radiatively-cooled media.    The question remains, however, if the transition we have found in our simplified models directly corresponds to a similar transition in the much more complex ISM distributions that occur in real galaxies.   Based on both theory and observations, we argue that there are many reasons to believe that it does. From a theoretical point of view, one can list a number of  ISM processes that we do not include in our simulations, yet none of them is likely to impede the turbulent runaway we see above  $\sigma_{\rm 1D, M} = 35$ km/s.  This runaway arises because any medium with sustained turbulence will be continuously heated by turbulent decay, and this thermal energy must be radiated away to maintain equilibrium.   This means that if turbulence is vigorous enough to move a fraction of the gas into a regime in which the cooling time increases strongly with temperature, the continued deposition of thermal energy will heat this gas dramatically, driving it out of the midplane, and in many cases out of the host galaxy itself.   The simplicity of this mechanism suggests that more complete ISM models will only refine the connection between this processes and galaxy outflows.

Thus while our nonrotating simulations do not include the Coreolis force, this is unlikely to play a key role.   In particular, its impact can be estimated by calculating the Rossby number, the ratio of the typical velocity to the shear across the region, which is greater than 1 in cases in which rotation is negligible. In a simulation offset from the center of the disk  by a distance $d_0,$ this is $R \approx \sigma_{\rm 3D,T d_0} / (2H v_0),$ where $v_0$ is the rotational velocity at $d_0$.   For a galaxy with a flat rotation curve $v_0^2 = g_0 d_0$ where $g_0$ is the gravitational acceleration as in eq\, (\ref{eq:g}), and for solid body rotation, as in a dwarf galaxy  or near the center of a large galaxy, $v_0^2 = 3 g_0 d_0$.    This means $R \approx (\sigma_{\rm 1D,T}/c_{\rm s,grav})Ê(d_0/H)^{1/2}$ for a flat rotation curve, and $R \approx (\sigma_{\rm 1D,T}/c_{\rm s,grav}) (3d_0/H)^{1/2}$ for solid body rotation,  where $c_{\rm s,grav} = (kT_{\rm grav}/\mu m_p)^{1/2} = 37 \, {\rm km/s} \, (T_{\rm grav}/10^5 K)^{1/2}.$   This means that $R \approx 1$  if we were to consider our lowest velocity simulation to take place at the center of a galaxy, and $R \gsim 1$ for all other simulations and choices of $d_0$.  Furthermore, the Coreolis force has no effect for velocities perpendicular to the disk, the direction of the gas ejection driven by turbulent runaway.  Rotation may be a mechanism worth exploring, but it is one that is unlikely to drastically effect the transition identified here.
 
Magnetic fields are also not included in our simulations, although they are known to exist in galaxies, and they can contribute to turbulence via the magentorotational instability (Piontek \& Ostriker 2004; 2005), modify the  flow dynamics, and affect the decay of turbulent velocities into heat.  On the other hand,  simulations of supersonic magnetohydrodynamic turbulence (Stone \etal 1998; Mac Low \etal 1998) have shown that the energy density of the magnetic fields built up in the MHD turbulence are typically $\approx 2-3$ smaller than the turbulent energy density, a result that is consistent with observations (Ferri\' ere 2001).  Furthermore, Lemaster \& Stone (2009) showed that large changes in magnetic field strength only have a minor impact on the dissipation rate of turbulence.   On the other hand, much like gravity, ordered fields have the potential to influence the flow near the plane of the galaxy,  even in the presence of large velocities and high temperatures.  Thus, magnetic fields are likely to lead to interesting changes in the simulations, but unlikely to affect the  dynamics strongly enough to avoid turbulent runaway at high velocity dispersions. 
 
While heat conduction has the potential to affect turbulent heating by moving energy from hot rarified regions into the cooler clumps, the addition of this processes to our simulations is very  unlikely to have any effect on our conclusions. Although the precise value of the conductivity is relatively uncertain and dependent on magnetic field structure (\eg Cowie \& McKee 1977; Malyshkin \& Kulsrud  2001), it is clear that it is a slow, second order process on ISM scales.   Instead,  mixing between hot and cold regions is primarily determined by the cascade of structures down to the very small scales at which conduction can operate (\eg Pan \& Scannapieco 2010).   In fact, if anything, as  turbulence is driven at marginally-resolved scales in our simulations, the presence of unavoidable numerical diffusion is likely to somewhat {\em overestimate} heat conduction and mixing relative to the real ISM, as evidenced in the increased mixing in our fiducial run, S34, relative to the high resolution run, S35HR.

Our simulations also do not include self-gravity effects, which can both serve to enhance shear-driven turbulence and lead to the formation of Jeans unstable molecular clouds.  While the swing amplifier instability that occurs in self-gravitating rotating disks (Goldreich \& Lynden-Benn 1965; Julian \& Toomre 1966)  has been shown to be important in low density regions near the edges of disk galaxies (\eg Wada, Meurer, \& Norman 2002; Kim \& Ostriker 2007; Koyama \& Ostriker 2009), it is unlikely to drive turbulence at the levels considered here.  On the other hand, molecular clouds can drive more vigorous turbulence by gravitational scattering (Fukunaga \& Tosa 1989; Gammie, Ostriker, \& Jog 1991), and much more importantly, by the formation of stars.    For any complete model of ISM evolution, tracking the formation of such clouds is essential, which requires not only self-gravity, but a full model of the molecular chemistry and cooling.   On the other hand, as our goal here is not to capture star formation in detail, but rather to explore the evolution of stratified media as a function of turbulent velocity dispersion, this process is only of secondary importance.  Our aim is not to capture how turbulence is driven, but rather study how driven turbulence affects the subsequent evolution of the ISM. 

Besides driving turbulence, star formation will also have two major effects  that are not considered in our simulations.   First it will lead to diffuse gas heating through the photoionization of neutral gas (\eg Wolfire et al 2003), and the photoelectric gas heating caused by the high-energy electrons that are removed from dust grains by stellar ultraviolet radiation (\eg Joung \& Mac Low 2006).  To the extent to which such heating varies with space and time (Parravano \etal 2003) or contributes to gas motions through radiation pressure (Thompson \etal 2005) it will contribute to the overall level of turbulence.   To the extent to which such heating is uniform, however, it will offset the overall level of cooling, which in turn will mean that less turbulent driving is needed to maintain equilibrium.  In this case, a given turbulent velocity dispersion will correspond to a larger overall column depth, again making the point that $H \bar n$ in our simulations is likely to be much less than measured in similar systems in nature, and that our results are much better interpreted as a function of $\sigma_{\rm 1D,M}.$ 

Secondly, supernovae and stellar winds arising from star formation will not only drive turbulence, but will also deposit hot gas directly into the interstellar medium.   In fact, the overlapping of such hot regions into a superbubble  is traditionally thought of as the mechanism by which outflows are driven (\eg Tomisaka \& Ikeuchi 1986; Mac Low \etal 1989;  Tenorio-Tagle \etal 1990; Mac Low \& Ferrara 1999; Strickland \& Stevens 2000;  Fujita \etal 2003).   On the other hand, there is considerable observational evidence supporting the idea that galaxy outflows are  driven instead by the more widespread input from the massive stars (Strickland \etal 2004).  For example, observed outflow pressure profiles demonstrate that mass and energy are ejected relatively smoothly over large regions  that match up well with region of intense star formation (Heckman \etal 1990; Lehnert \& Heckman 1996; Strickland \& Stevens 2000; Strickland \etal 2000).  

This suggests that turbulent runaway may in fact be the primary mechanism producing the hot gas observed to be pouring out of starbursting galaxies, and that the fraction of hot gas directly added by SNe is not only uncertain theoretically, but of secondary importance physically.  In fact, adding this process to our simulations is only likely to obscure the main results.   The ISM in galaxies with velocity dispersions below $\sigma_{\rm 1D,M}$ would contain hot gas, but this gas would cool and condense within the disk relatively quickly. On the other hand, supernova material in $\sigma_{\rm 1D,M}$ disks would either be mixed into the ISM if it were deposited within a dense clump, or escape to high latitudes if it were ejected into a low-density, superheated region.   As seen in our more approximate galaxy-scale modeling in Scannapieco \& Br\"uggen (2010), this would lead to outflows that were more metal enriched than the overall ISM, but this difference would be much less than in a ``superbubble" outflow picture in which the metals from supernovae were injected into a single hot cavity.  However, as our simulations include energy input from stars as random turbulent motions its is not possible for us to associate metals from stars with particular parcels of gas in our present simulations to quantify this effect further.  Furthermore, the best way to implement supernovae into ISM scale simulations is extremely unclear, and to date has lead to turbulent velocity dispersions that vary only weakly with star formation rate and increase only up to $\approx 15 $ km/s, (Dib \etal 2006; Shetty \& Ostriker 2008; Agertz \etal 2009; Joung \etal 2009), falling far short of the critical velocity dispersion identified here.

On the other hand, with the advent of modern integral-field unit (IFU) spectroscopy, galaxies with turbulent velocities above 35 km/s have been well observed at a variety of redshifts, with a variety of methods.  At cosmological distances, Lemoine-Busserolle \etal (2010) and Lemoine-Busserolle \& Lamareille (2010) conducted  Adaptive Optics (AO) IFU spectroscopy with the VLT/SINFONI instrument, using rest-frame optical nebular emission lines to measure average velocity dispersions of $60-70$ km/s in four $z \approx 3$ galaxies and velocity dispersions of $35-100$ km/s in eight $z= 1.0- 1.5$ galaxies.  At Keck, Jones \etal (2010) conducted AO OSIRIS IFU spectroscopic observations of six strongly-lensed $z = 1.7-3.1$ galaxies, at resolutions of up to 100 pc, obtaining H$\alpha$ velocity dispersions between $\sigma_{\rm 1D, H\alpha} = 50-100$ km/s.  Wright \etal (2007) and Law \etal (2009) used AO and the  OSIRIS IFU  to study a sample of unlensed  galaxies between $z=1.5$ and $3.1$ and found $\sigma_{\rm 1D, H\alpha}$  between $50-100$ km/s in most cases, and $139 \pm 3.2$ in an extreme starbursting region of a single galaxy (Q1623-BX543).  van Starkenburg \etal (2008)  used SINFONI at the VLT to observe a $z=2.03$ galaxy with  $\sigma_{\rm 1D, H\alpha}$ ranging between $30-100$ km/s, and Epinat \etal (2009) used this instrument to observe nine  $z = 1.2 - 1.6$ galaxies with $\sigma_{\rm 1D, H\alpha} \approx 40 - 100$ km/s.  Finally, the large SINS survey  to study  the dynamics of high-redshift galaxies also used SINFONI IFU spectroscopy,  and to date has measured over twenty $z \approx 2$ disk galaxies and star-forming clumps with $\sigma_{\rm 1D, H\alpha} = 30-100$ km/s (F\" orster Schreiber \etal 2006; 2009; Cresci 2009; Genzel \etal 2008; 2011).

In the nearby universe, the Gassendi H$\alpha$ survey of SPirals (GHASP) has used spatially-resolved Fabry-P\' erot spectroscopy to observe a  sample of 153 isolated disk galaxies with $H\alpha$ velocity dispersions  ranging from 20-40 km/s (Epinat \etal 2008; 2010).  Selecting a sample of higher H$\alpha$ luminosity disks from the Sloan Digital Sky Survey, Green \etal (2010) used  the SPIRAL and WiFeS IFU spectrographs to measure luminosity-weighted  $H\alpha$ velocity dispersions on $\approx 2$ kpc scales in a sample of  65 $z \approx 0.1$ rapidly star-forming galaxies.   These velocity dispersions ranged from $25-100$ km/s, and they were strongly correlated with star formation rate and uncorrelated with stellar mass, suggesting massive stars as the primary drivers of vigorous turbulence.   

In observed star forming disk galaxies, the total pressure in the disk is just sufficient to stabilize all perturbations on scales too small to be stabilized by rotation, meaning that the Toomre parameter $Q = (\sigma^2_{\rm 1D}+ c_s^2)^{1/2} \kappa/ \pi G  \Sigma_g \approx 1-2,$ where $\kappa$ is the epicyclic frequency and $\Sigma_g$ is the gas density per unit area (Leroy \etal 2008).  This means that the turbulent velocity dispersion is closely tied up with the gas surface density, which in turn is closely tied up with the star formation rate per unit area $\dot \Sigma_\star.$  In fact Genzel \etal (2011), compiled a large sample of the higher-redshift data described above to show that all galaxies with $\sigma_{\rm 1D,H\alpha} \geq 35$ km/s have high star formation rate densities $\dot \Sigma_\star > 0.1 M_\odot \, {\rm yr}^{-1} \, {\rm kpc}^{-2}.$   At the same time, outflows are ubiquitous in all  galaxies in which the star formation rate per unit area exceeds this same value of $0.1 M_\odot \, {\rm yr}^{-1} \, {\rm kpc}^{-2}$ (Heckman 2002).  Indeed, the star forming nucleus in M82, the archetypal example of an outflowing starburst, has a star formation surface density of $\approx 5 M_\odot \, {\rm yr}^{-1} \, {\rm kpc}^{-2}$ (Strickland \& Heckman 2009) and an average velocity dispersion of  $\approx 90 \pm 30$ km/s (Westmoquette \etal 2009).  We may be just beginning to uncover the role of turbulent runaway in the history of galaxy formation.

\acknowledgements

We are grateful to Robert Fisher, Mark Krumholz, Crystal Martin, Prateek Sharma, and Natascha F\" orster Schreiber  for their helpful comments. The ideas described here grew out of discussions at the Star Formation in Galaxies Summer Workshop at the Aspen Center for Physics. Parts of this work were performed while ES attended the Galaxy Clusters: the Crossroads of Astrophysics and Cosmology Workshop at the Kavli Institute for Theoretical Physics at UCSB. We acknowledge support from NASA under theory grant NNX09AD106 and the National Science Foundation under grants AST 11-03608 and PHY05-51164. All simulations were conducted at the ASU Advanced Computing Center, using the FLASH code, a product of the DOE ASC/Alliances-funded Center for Astrophysical Thermonuclear Flashes at the University of Chicago.

\end{document}